# *ATRIA*: A Bit-Parallel Stochastic Arithmetic Based Accelerator for In-DRAM CNN Processing

Supreeth Mysore Shivanandamurthy, Ishan. G. Thakkar, Sayed Ahmad Salehi
*Department of Electrical and Computer Engineering, University of Kentucky, Lexington, KY,USA.*
*supreethms@uky.edu, igthakkar@uky.edu, sayedsalehi@uky.edu*.

***Abstract*— With the rapidly growing use of Convolutional Neural Networks (CNNs) in real-world applications related to machine learning and Artificial Intelligence (AI), several hardware accelerator designs for CNN inference and training have been proposed recently. In this paper, we present *ATRIA*, a novel bit-pArallel sTochastic aRithmetic based In-DRAM Accelerator for energy-efficient and high-speed inference of CNNs. *ATRIA* employs light-weight modifications in DRAM cell arrays to implement bit-parallel stochastic arithmetic based acceleration of multiply-accumulate (MAC) operations inside DRAM. *ATRIA* significantly improves the latency, throughput, and efficiency of processing CNN inferences by performing 16 MAC operations in only five consecutive memory operation cycles. We mapped the inference tasks of four benchmark CNNs on *ATRIA* to compare its performance with five state-of-the-art in-DRAM CNN accelerators from prior work. The results of our analysis show that *ATRIA* exhibits only 3.5% drop in CNN inference accuracy and still achieves improvements of up to 3.2× in frames-per-second (FPS) and up to 10× in efficiency (FPS/W/mm$^2$), compared to the best-performing in-DRAM accelerator from prior work.**

## I. INTRODUCTION

Convolutional Neural Networks (CNNs) have achieved remarkable progress in recent years, and they are being aggressively utilized in real-world applications related to Artificial Intelligence (AI) and machine learning [9][10]. In general, CNNs mimic biological neural networks, and utilize compute-heavy arithmetic functions such as multiply-accumulate (MAC), nonlinear activation, and pooling. Although these CNN functions are amenable to acceleration because of a high degree of compute parallelism, their acceleration using traditional ASIC platforms (e.g., Dadiannao [9], EIE [24]) is challenging because of the need to avoid the memory wall while accessing their large number of operands [12]. To address this problem, several prior works have explored processing-in-memory (PIM) designs based on the emerging non-volatile memory (NVM) crossbar technologies (e.g., ISAAC [10], PRIME [22], XNOR-RRAM [23]) as well as the traditional DRAM technology (e.g., DRISA [1], SCOPE [2], DRACC [21], LACC [3]). Such PIM designs strive to avoid data movement to consequently achieve a balance between computational efficiency and memory performance while processing CNNs in situ.

However, it is challenging to support MAC operations in PIM designs. The NVM crossbar-based PIM designs, such as ISAAC [10] and PRIME [22], leverage the Kirchoff's Law to perform MAC operations in the analog domain. However, such analog computing-based accelerators require power-hungry and throughput-limited digital-to-analog converters (DACs) and analog-to-digital converters (ADCs), which diminishes the performance and energy-efficiency benefits of such accelerators. Alternatively, the DRAM-based PIM designs implement in-situ MAC operations digitally, for which they break a single MAC operation into multiple functionally complete memory operation cycles (MOCs) that are serially run on a single subarray (the smallest logical cell array in a DRAM module). Multiple such subarrays typically work in parallel to achieve high processing throughput. Such designs require a very larger number of MOCs per MAC operation. For instance, DRISA [1] requires up to 222 MOCs per MAC. To reduce the required number of MOCs, SCOPE [2], DRACC [21], and LACC [3] employ light-weight optimizations that simplify the in-DRAM implementation of MAC operations. SCOPE adopts stochastic arithmetic to implement approximate multiplication, requiring a reduced number of up to 25 MOCs per MAC [2]. In contrast, DRACC [21] eliminates most multiply operations by employing quantized CNNs that use ternary weights, whereas LACC [3] employs lookup table based multiply operations. Because of these optimizations, DRACC and LACC require reduced number of MOCs per MAC of up to 13 and 11 respectively. This can still incur very high latency and energy consumption as one MOC can incur up to 49 ns latency and up to 4nJ energy consumption [1][11][21], depending on the utilized DRAM technology node, and subarray size (bitline length). The high latency and energy values per MAC operation have prevented the DRAM-based PIM designs from being immediately adopted for CNN inference.

In this paper, we present a novel CNN accelerator called *ATRIA*. *ATRIA* employs bit-parallel stochastic arithmetic, which enables it to perform 16 MAC operations in only 5 consecutive MOCs. *ATRIA* is most related to SCOPE [2]. It significantly improves upon SCOPE in *two* ways. *First*, SCOPE uses stochastic arithmetic to perform only multiply operations, whereas it uses the conventional binary arithmetic to perform accumulate operations. In contrast, *ATRIA* performs both multiply and accumulate operations using bit-parallel stochastic arithmetic. *Second*, both SCOPE and *ATRIA* require expensive binary-to-stochastic (B-to-S) and stochastic-to-binary (S-to-B) conversions of operands, but *ATRIA* is better able to hide the latency of these conversions by successfully removing them from the critical processing path. Moreover, *ATRIA* restricts the precision errors induced due to the stochastic arithmetic based accumulate operations by employing stochastic operands that are 2× larger than their full-precision size. As a result, *ATRIA* exhibits only 3.5% drop in CNN inference accuracy on average compared to SCOPE. Despite this slight drawback, *ATRIA* substantially outperforms SCOPE as well as other in-DRAM accelerators, such as DRISA and LACC, in terms of the latency, throughput (frames-per-second (FPS)), and efficiency (FPS/W/mm$^2$) of processing state-of-the-art CNNs.

Our key contributions in this paper are summarized below.

- We present a novel accelerator architecture called *ATRIA* that is designed for in-DRAM processing of CNN inference tasks;
- We introduce a novel concept of bit-parallel stochastic arithmetic, which enables *ATRIA* to perform 16 MAC operations in only five consecutive memory operation cycles;
- We employ low-overhead add-on logic in *ATRIA* to implement binary arithmetic-based activation and pooling functions directly inside DRAM subarrays;
- We evaluate *ATRIA* for four state-of-the-art deep CNN topologies, i.e., VGG16 [14], Alexnet [13], ResNET_50 [16], GoogleNET [15], with the *ImageNet* dataset;
- We compare the performance of *ATRIA* for the considered CNNs with the following five in-DRAM CNN accelerators from prior work: DRISA-3T1C [1], DRISA-1T1C-NOR [1], SCOPE-Vanilla [2], SCOPE-H2D [2], and LACC [3].

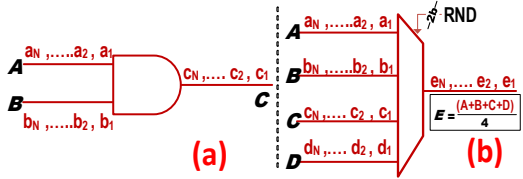

Fig. 1. Bit-serial stochastic arithmetic circuits for (a) multiplication (AND gate), (b) scaled accumulation (MUX).

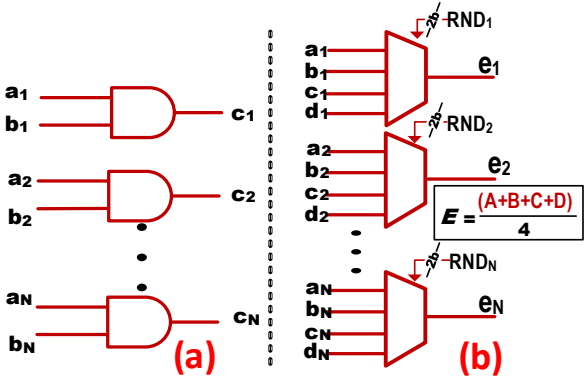

Fig. 2. Bit-parallel stochastic arithmetic circuits for (a) multiplication (array of AND gates), (b) scaled accumulation (array of MUXs). Here, the individual N bits of operands A, B, C, and D from Fig. 1 are striped across N copies of AND gates and MUXs.

## II. CONCEPT OF BIT-PARALLEL STOCHASTIC ARITHMETIC

The use of stochastic arithmetic simplifies the implementation of complex arithmetic functions, such as multiplication and accumulation, by reducing them to simple bit-wise logical operations [4]. To perform a multiplication of 2 N-bit stochastic operands (A and B in Fig. 1(a)) in the bit-serial manner, the bit-streams of the operands are applied to an AND gate serially, and the bit-wise output of the AND gate is collected for total N clock cycles to generate the multiplication output bit-stream (C in Fig. 1(a)). Similarly, to perform a scaled accumulation of 4 (or more) N-bit stochastic operands in the bit-serial manner (A, B, C, D in Fig. 1(b)), the bit-streams of the operands are applied to a MUX, whose bit-wise output is selected by a 2-bit (or larger) random number (RND in Fig. 1(b)) every clock cycle for total N clock cycles, to generate the output bit-stream that represents a scaled accumulation (E in Fig. 1(b)). To reduce the area and static power consumption of computing, such bit-serial implementation of stochastic arithmetic compromises the latency of computing.

In contrast, we observe that the latency of computing can be improved by N× if the stochastic arithmetic can be implemented in the bit-parallel manner. For example, if N copies of AND gates and MUX circuits are available (Figs. 2(a) and 2(b)), the N-bit outputs for the stochastic multiplication and scaled accumulation can be obtained in one clock cycle in the bit-parallel manner. In a nutshell, the idea for such bit-parallel implementation of stochastic arithmetic is to transform the input bit-streams into bit-vectors by striping them across the N copies of the AND gates or MUX circuits, and then perform bit-wise AND or MUX operations to generate output bit-vectors. For instance, the individual N bits $a_1$ to $a_N$, $b_1$ to $b_N$, $c_1$ to $c_N$, and $d_1$ and $d_N$ of operands A, B, C, and D from Fig. 1(b) are striped across N copies of MUXs in Fig. 2(b). Similarly, the individual N bits $a_1$ to $a_N$ and $b_1$ to $b_N$ of operands A and B from Fig. 1(a) are striped across N copies of AND gate in Fig. 2(a). As a result, the individual N bits of the scaled accumulation output E in Fig. 2(b) (or scaled multiplication output C in Fig. 2(a)) can be collected in a bit-parallel manner from N MUXs (or N AND gates). For such bit-parallel scaled accumulation, total N RND signals ($RND_1$ to $RND_N$) are needed to select the output of the MUXs, which can be generated a priori and made available in a parallel manner (Fig. 2(b)). Although Fig. 2(b) illustrates bit-parallel scaled accumulation for only four input stochastic operands (A, B, C, and D), this concept can be extended for more or less than 4 stochastic operands as well.

Such bit-parallel stochastic arithmetic naturally fits well for in-DRAM processing of applications because the inherent parallelism of DRAM makes it fundamentally easy to provision data in the bit-parallel manner. *Our proposed in-DRAM accelerator ATRIA is the first to employ such bit-parallel stochastic arithmetic for implementing in-DRAM MAC operations. Exploiting the benefits of such implementation, ATRIA substantially improves the latency and throughput of in-DRAM CNN processing with only a marginal decrease in CNN inference accuracy, compared to the in-DRAM CNN processing accelerators from prior work.*

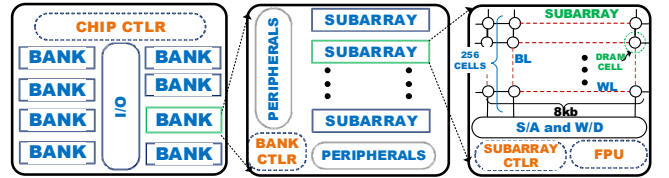

Fig. 3. The hierarchical structure of our *ATRIA* accelerator chip.

## III. *ATRIA*: OVERVIEW

Our *ATRIA* accelerator architecture employs an 8Gb DRAM module with 8 chips. Fig. 3 illustrates the schematic of one such chip. Each chip has 8 banks, with 64 subarrays per bank, and 32 mats per subarray of 256×256 bits size each. Each row in a subarray is of 8Kb size, therefore, each subarray contains total 8Kb sense amplifiers (S/As) and write drivers (W/Ds). Each subarray acts as a processing element (PE), which is defined as the smallest independent cell-array structure that can perform computing. Therefore, there are total 4096 PEs in *ATRIA*. Like the other in-DRAM accelerators from prior work (e.g., DRISA [1], SCOPE [2], LACC [3]), the PEs in *ATRIA* can also operate in parallel to process CNN inference in situ. To process CNN inference, each PE (i.e., subarray) in *ATRIA* employs a feature processing unit (FPU), as shown in Fig. 3. In addition, to orchestrate these in-parallel processing operations inside the PEs, *ATRIA* employs hierarchical controllers (chip, bank and subarray controllers (CTLRs) in Fig. 3). The operation of these hierarchical CTLRs is described in Section III.C. The structure and operation of each FPU in *ATRIA* support our concept of bit-parallel stochastic arithmetic for in-situ processing of CNNs, as discussed next.

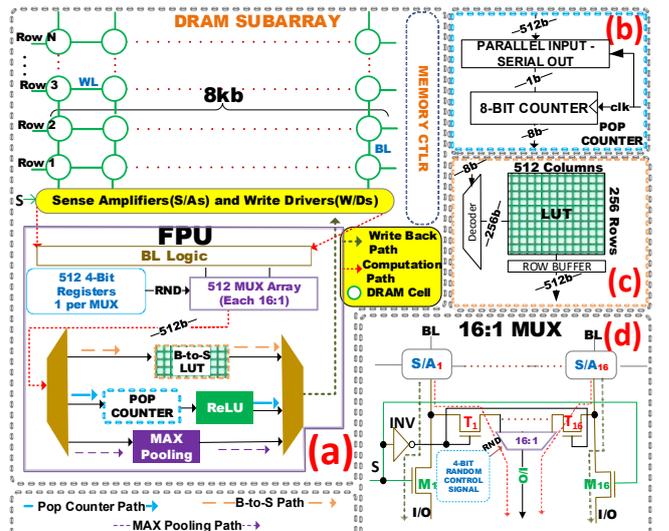

Fig. 4. Schematic of a processing element (PE) of *ATRIA*. (a) Schematic of a subarray and feature processing unit (FPU); (b) pop counter for S-to-B conversion; (c) LUT for B-to-S conversion [2]; (d) a 16:1 MUX and its connections with S/As as part of the FPU.

## A. Structure of a PE in ATRIA

A PE of our *ATRIA* accelerator is basically a DRAM subarray that is integrated with an FPU and a subarray CTLR, as illustrated in Fig. 4. The subarray part of the PE is structured in the manner the conventional DRAM subarrays are organized [5][28]. Therefore, in this section, we only provide details of the structure of the FPU. The role of the subarray CTLR is discussed in Section III.C. The FPU consists of various hardware components that support the implementation of the following *six* functions: (i) bit-parallel stochastic multiply operation (MUL), (ii) bit-parallel stochastic accumulate operation (ACC), (iii) binary to stochastic (B-to-S) conversion, (iv) stochastic to binary (S-to-B) conversion through pop counter (PC), (v) nonlinear activation function ReLU, and (vi) max pooling function. To support bit-parallel MUL, three 8Kb rows of the subarray (Row 1, Row 2 and Row 3 in Fig. 4(a)) are reserved and operated following the triple row activation and charge-sharing protocol of AAP memory operation cycle (MOC) from Ambit [11] (see Section III.B).

The hardware components that support bit-parallel ACC consist of an array of 512 copies of 16:1 MUXs and their associated 512 copies of 4-bit registers (Fig. 4(a)). These 4-bit registers store the pre-determined random values that enable the output selection (16:1) for their respective MUXs. Each MUX has 16 inputs, therefore, the total number of inputs for the entire array of 512 MUXs is 8Kb. These 8Kb MUX inputs are connected to 8Kb S/As, with 16 adjacent S/As feeding one MUX and vice versa (Fig. 4(d)). Note that the S/As in the commodity DRAMs typically connect to I/O logic through signal S and related control transistors ($M_1$ to $M_{16}$) (Fig. 4(d)). To facilitate connections of S/As to MUXs, *ATRIA* additionally employs one inverter (INV) and 16 transistor switches ($T_1$ to $T_{16}$) per MUX, which can be controlled by the same signal S (Fig. 4(d)). An 8Kb row from the subarray can be read into 8Kb S/As (Fig. 4(a)), which can hold total 16 stochastic bit-vectors of 512-bit size each (16 × 512 = 8Kb). These 16 stochastic bit-vectors can be striped across 512 MUXs, so that each individual bit of a bit-vector is fed into a different MUX with each MUX having all its 16 inputs from 16 different bit-vectors. This arrangement sets up the array of MUXs to perform a 16-operand scaled ACC in the bit-parallel manner, following our concept of bit-parallel stochastic arithmetic discussed in Section II. The detailed functioning of this array of MUXs for performing scaled ACC is presented in Section III.B.

In addition, to implement in-memory B-to-S conversion, each FPU in *ATRIA* employs a lookup table (Fig. 4(c)). Our idea of using lookup table-based B-to-S conversion is inspired from the design of SCOPE accelerator [2]. This enables *ATRIA* to employ the deterministic method for B-to-S conversion to eliminate correlation errors [2]. Moreover, each FPU in *ATRIA* employs an additional lookup table to perform ReLU (Fig. 4(a)). Further, it also incorporates a pop counter to perform in-memory S-to-B conversion (Fig. 4(b)), as well as logic to implement max pooling function (Fig. 4(a)). *ATRIA* implements the max pooling and ReLU functions in the binary domain. This mandates that the results of processing of every CNN layer's parameters always go through S-to-B, ReLU, and then B-to-S conversions before they can activate processing of the next CNN layer. This in turn eliminates the undesirable propagation of precision errors (which are very common in stochastic arithmetic [4]) between the stochastic operations of two consecutive CNN layers (see more on errors in Section IV.B). The overheads of incorporating FPUs in *ATRIA* PEs are discussed in Section III.D. The next section describes the functioning of an FPU-enabled PE of our *ATRIA* accelerator.

## B. Functioning of a PE in ATRIA

Each PE of our *ATRIA* accelerator can perform all essential functions required for processing CNNs, such as MAC, max pooling, and ReLU. In addition, since *ATRIA* employs stochastic arithmetic, each PE can also perform important functions for implementing stochastic arithmetic, such as B-to-S and S-to-B (pop count) conversions. On one hand, each PE performs B-to-S, S-to-B (pop count), ReLU, and max pooling functions by relaying the related operands along the data processing path in the FPU through the corresponding hardware components (Fig. 4(a)). To orchestrate the relaying of the operands to perform these functions, the PE makes use of the subarray CTLR whose functioning along with the functioning of other hierarchical CTLRs in *ATRIA* is discussed in Section III.C. On the other hand, each PE of *ATRIA* can perform a MAC function ($F_{MAC}$) of 16 stochastic operands of 512-bit size each, by employing a series of total five memory operation cycles (MOCs) (similar to the AAP/AP MOC from [1][11]). These MOCs engage the reserved rows Row 1, Row 2 and Row 3 (Fig. 4(a)) and the MUXs in the FPU, as discussed next.

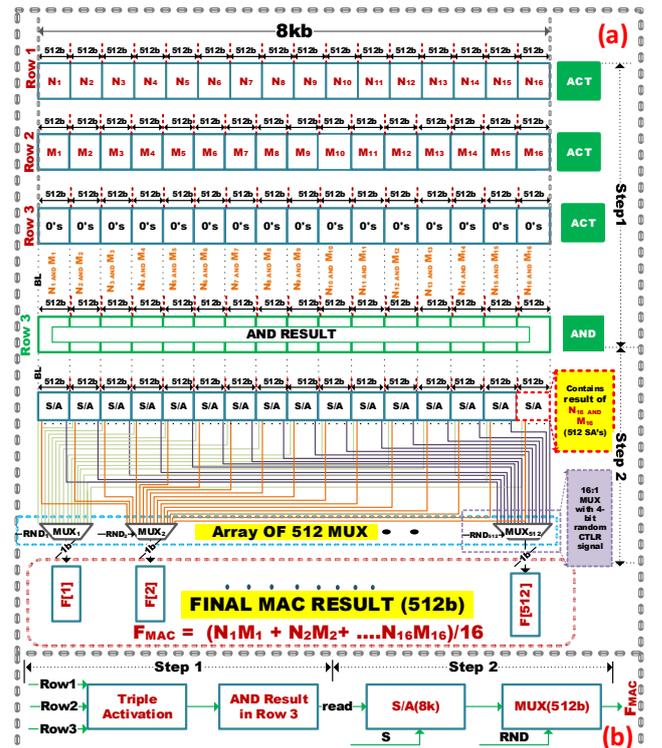

**Fig. 5.** A schematic showing the operation of a PE of *ATRIA* to perform a 16-operand multiply-accumulate (MAC) function ($F_{MAC}$).

Fig. 5 illustrates how *ATRIA* performs $F_{MAC}$. *ATRIA* performs $F_{MAC}$ in two main steps. Step 1 engages the reserved subarray rows Row 1, Row 2, and Row 3 to perform MUL. Step 2 engages the array of MUXs to perform ACC. Before performing $F_{MAC}$, *ATRIA* first makes the involved stochastic operands available in the reserved subarray rows Row 1 and Row 2. For that, it performs two MOCs similar to RowClone [25] to copy the contents of two source rows into Row 1 and Row 2 respectively. Consequently, Row 1 contains 16 512-bit operands $N_1$ to $N_{16}$ (Fig. 5(a)). Similarly, Row 2 contains 16 512-bit operands $M_1$ to $M_{16}$ (Fig. 5(a)). In addition, *ATRIA* initializes Row 3 with '0's at system boot. After these initial steps, *ATRIA* schedules Step 1 of $F_{MAC}$, which employs the triple row activation and charge-sharing based MOC from Ambit [11] to perform bit-parallel logical AND (i.e., stochastic MUL) of the involved operands $N_1$ to $N_{16}$ and $M_1$ to $M_{16}$. At the end of the MOC for Step 1, Row 3 contains the results of bit-parallel logical AND, i.e., $N_1$ AND $M_1$ to $N_{16}$ AND $M_{16}$ (Figs. 5(a) and 5(b)). These results essentially represent the outcome of bit-parallel stochastic MUL, i.e., $N_1M_1$ to $N_{16}M_{16}$. After this, *ATRIA* schedules Step 2 of $F_{MAC}$, where it performs a MOC to read the stochastic MUL results from Row 3 into S/As. These results from S/As are then pushed through the array of 16:1 MUXs, $MUX_1$ to $MUX_{512}$. The 512-bit output of this array of MUXs is selected using the pre-latched random control signals $RND_1$ to $RND_{512}$. This 512-bit output is the stochastic scaled ACC of the input operands $N_1M_1$ to $N_{16}M_{16}$. In other words, this 512-bit output presents $F_{MAC} = (N_1M_1 + N_2M_2 + \ldots + N_{16}M_{16})/16$ (Figs. 5(a) and 5(b)). *ATRIA* then uses one more MOC to store the result of this $F_{MAC}$ into a row in the subarray through W/Ds. Thus, *ATRIA* uses only 5 MOCs (2 MOCs for initializing Row 1 and Row 2, 1 MOC for MUL, 1 MOC for ACC, and 1 MOC for write back) to perform a scaled MAC function $F_{MAC}$ (also called dot-product) of 16 stochastic operands. In other words, if a MAC operation is conventionally defined as a MUL of two operands followed by an accumulate operation (i.e., $A = A + N_iM_i$), then *ATRIA* uses only 5 MOCs to perform 16 MAC operations in parallel. However, we find from our evaluation results in Section IV.B that the use of bit-parallel stochastic arithmetic in *ATRIA* can increase precision errors. Nevertheless, we also find that the increased precision errors are worth tolerating for due to the substantial performance benefits of *ATRIA*.

### C. System Integration and Controller Design

In this section, we describe how our *ATRIA* accelerator integrates with the host system and how the hierarchical controllers of *ATRIA* orchestrate the processing of CNNs. *ATRIA* integrates with the host system in the same way the conventional GPU or FPGA based accelerators do through PCIe bus. For a CNN processing using *ATRIA*, the host system stores the weighting parameters and inputs of the CNN in the individual PEs (subarrays) of *ATRIA* via direct memory access (DMA). We adopt the strategy from SCOPE [2], wherein the weighting parameters are stored in *ATRIA* in the stochastic format a priori. This strategy ensures that in-situ B-to-S conversions are required only for inter-layer activation parameters, which significantly reduces the number of in-situ B-to-S conversions. As a result, *ATRIA* achieves dramatically reduced latency and energy of processing CNNs.

After storing the inputs and weighting parameters of a CNN in PEs of *ATRIA*, the host-side *ATRIA* CTLR (not shown in Fig. 3) orchestrates the processing of the CNN in conjunction with the hierarchical *ATRIA* CTLRs shown in Fig. 3. The host-side *ATRIA* CTLR generates a series of μ-operations, which are received by the hierarchical *ATRIA* CTLRs. We adopt the designs from [1] for these CTLRs. These CTLRs support simultaneous multi-subarray/bank activation for better parallelism. The first chip-level CTLR is essentially a decoder, and it also helps with inter-bank data movement. The bank-level CTLRs decode the μ-operations and convert them into addresses, vector lengths, and control codes, and then send them to subarray CTLRs in the active subarrays. The subarray CTLR consists of address latches, local decoders, and counters. The address latches are essential for multi-subarray activation [1]. The counters are used for continuously updating addresses to local subarray decoders. In addition, the subarray CTLR also contains buffers to support communication of operands.

Inter-bank and inter-subarray data communications in *ATRIA* are supported through the interconnects design adopted from LISA [26]. Data communications are carried out in binary format instead of stochastic format, which results in better energy-efficiency [2]. Also, the inclusion of buffers in the subarray CTLRs enables pipelined data communications, which enables better use of resources and efficient hiding of long latencies, reducing the memory bottleneck to improve the throughput of CNN processing with *ATRIA*.

### D. Overhead Analysis

Table 1 lists the latency, energy, and area overheads of various hardware components that are part of the FPUs inside the PEs of our *ATRIA* accelerator. These results are based on our logic synthesis analysis for 22nm node. We considered standard SRAM for LUT implementation. After accounting for the extra area overhead of these components from Table 1, the total area for 8Gb *ATRIA* accelerator becomes $77mm^2$. In comparison, DRISA-1T1C-NOR [1], DRISA-3T1C [2], SCOPE-Vanilla [2], SCOPE-H2D [2], and LACC [3] consume $55mm^2$, $64.6mm^2$, $259.4mm^2$, $273.4mm^2$, and $61mm^2$ area respectively. Thus, *ATRIA* consumes larger area than DRISA-1T1C-NOR, DRISA-3T1C, and LACC. Nevertheless, *ATRIA* still achieves substantially better area and energy efficiency compared to these accelerators (Section IV.D). Similarly, despite the S-to-B pop counter in *ATRIA* incurring a long latency of 256ns (Table 1), the performance of *ATRIA* does not get much affected, as *ATRIA* manages to keep this latency out of the critical processing path (Section IV.C).

**Table 1. Latency, energy, and area overhead values of various hardware components of the FPUs in the PEs of *ATRIA*.**

| Component | Total Area ($mm^2$) | Latency (ns) | Energy per PE (pJ) |
|---|---|---|---|
| 16:1 MUXs for ACC | $1.3 \times 10^{-3}$ | 2 | 10 |
| 4-bit Registers for RND Storage | $1.1 \times 10^{-5}$ | 2 | 15.6 |
| B-to-S LUT (512×256) | 3.4 | 1 | 0.3 |
| S-to-B Pop Counter (PC) (2GHz) | $2.1 \times 10^{-5}$ | 256 | 153.6 |
| ReLU LUT | 1.2 | 1 | 0.3 |
| Max Pooling Logic | 4.1 | 5 | 940 |

## IV. EVALUATION

### A. Modeling and Setup for Evaluation

We evaluate *ATRIA* and compare it with other in-DRAM accelerators from prior work such as SCOPE-Vanilla [2], SCOPE-H2D [2], DRISA-1T1C-NOR [1], DRISA-3T1C [1], and LACC [3]. We first evaluate the per-MAC latency, per-MAC energy, and total area values for our considered accelerators. We divide the evaluation of per-MAC latency/energy into two parts: latency/energy of a multiply operation (MUL) and latency/energy of an accumulate operation (ACC). All our considered accelerators follow the AAP/AP memory operation cycle (MOC) from Ambit [11]. Therefore, the latency and energy values per MOC and total number of MOCs per MAC are evaluated first for all considered accelerators. Different accelerators have different latency and energy per MOC because they employ different lengths of local bitlines in their subarrays. For example, DRISA [1] and SCOPE [2] employ shorter local bitlines with only 64 cells per bitline. In contrast, LACC employs 512 cells per bitline, whereas *ATRIA* employs 256 cells per bitline. Shorter bitlines typically yield lower latency per MOC [5]. We evaluate latency using SPICE [18] based modeling of local bitlines. To evaluate per-MOC energy as well as total accelerator area, we used CACTI [17]. We developed a custom simulator in Python to model the MOC-accurate transaction-level performance behavior of our considered accelerators, as well as to evaluate system-level performance metrics such as frames-per-second (FPS), latency, efficiency (FPS/W/$mm^2$), and memory bottleneck ratio. Memory bottleneck ratio is defined as the ratio of total stall time (time for which an accelerator needs to wait for the operands) over total inference processing time. We considered four state-of-the-art CNNs to evaluate these metrics. The quantized versions of these CNN models were trained using PyTorch for ImageNet dataset and 8-bit fixed-precision of activation and weight parameters. These activation and weight parameters were extracted and provided as the input to our Python based performance simulator, which also took our evaluated energy, latency, and area values for our considered accelerators as the input. Next, we present and discuss the results of our simulation-based study.

## B. Precision Error and Accuracy Results

*ATRIA* has one caveat compared to SCOPE. The use of MUX based bit-parallel stochastic accumulation in *ATRIA* can increase the absolute precision error (APE) of computing, as explained in [4]. An APE for an operation (i.e., MUL or ACC) is defined as the absolute difference between the expected result and the observed result of the operation. From [4] and [27], APE depends on the operand values, input size (i.e., number of operands), and operand size (i.e., bit-stream length). For a MUX based stochastic ACC with an input size of 16 (as is the case for *ATRIA*), the average APE (μAPE) can be reduced to an acceptable value in the range between 0.2 to 0.54, if the operand size is kept 512 bits or longer [4][27]. Therefore, we increase the operand size, i.e., bit-vector length, of the bit-parallel stochastic operands in *ATRIA* to 512 bits from their full-precision length of 256 bits (corresponds to 8-bit binary operands). The resultant μAPE values and corresponding standard deviation in APE (σAPE) for four benchmark CNNs are listed in Table 2. The μAPE and σAPE values in Table 2 were obtained for the complete set of individual APEs for all MAC results required in respective CNNs when the inferences of these CNNs are implemented on *ATRIA*, SCOPE-Vanilla and SCOPE-H2D for the ImageNet dataset. Table 2 also lists the inference accuracy results. As evident, *ATRIA* exhibits 2.9× and 1.5× more μAPE, and 3.2× and 1.6× more σAPE than SCOPE-H2D and SCOPE-Vanilla respectively, on average across the CNNs. Nevertheless, compared to SCOPE-H2D and SCOPE-Vanilla, *ATRIA* exhibits only 3.5% and 0.85% drop in inference accuracy on average across the CNNs, which we reason is acceptable due to the significant performance benefits of *ATRIA*, as evident from Sections IV.C and IV.D.

**Table 2. Average APE (μAPE), standard deviation in APE (σAPE) and CNN testing accuracy (A) for SCOPE-Vanilla, SCOPE-H2D and *ATRIA* for various CNNs.**

| CNN Benchmarks | SCOPE-Vanilla | | | SCOPE-H2D | | | *ATRIA* | | |
|---|---|---|---|---|---|---|---|---|---|
| | μAPE | σAPE | A(%) | μAPE | σAPE | A (%) | μAPE | σAPE | A(%) |
| Alexnet [13] | 0.23 | 0.04 | 93.6 | 0.09 | 0.01 | 96.7 | 0.33 | 0.05 | 92.2 |
| GoogleNet [15] | 0.30 | 0.05 | 87.7 | 0.17 | 0.03 | 88.5 | 0.41 | 0.07 | 87.7 |
| VGG16 [14] | 0.35 | 0.05 | 91.9 | 0.21 | 0.03 | 95.1 | 0.53 | 0.09 | 90.2 |
| ResNet-50 [16] | 0.26 | 0.04 | 90.1 | 0.12 | 0.02 | 93.6 | 0.47 | 0.08 | 89.8 |

## C. Per-MAC Latency Results

Table 3 lists our evaluated latency values and number of PEs (#PEs) for *ATRIA* and other in-DRAM CNN accelerators. The latency values include values for MUL and ACC in number of MOCs (#MOCs), latency per MOC in ns, as well as the latency values for LUT-based B-to-S conversion and pop-count (PC) operations (required for S-to-B conversion). From Table 3, *ATRIA* holds *three crucial advantages*. *First*, it exhibits smaller per-MAC latency over SCOPE, DRISA and LACC (Table 3). This is because *ATRIA* performs 16 MAC operations in parallel. For that, *ATRIA* uses total 5 MOCs (total 85ns latency with each MOC incurring 17ns latency) (Section III.B); 2 MOCs to copy the operand rows, 1 MOC to perform 16 in-parallel MULs, 1 MOC to perform 16 in-parallel ACCs, and 1 MOC to store the MAC result. In Table 3, for *ATRIA*, 2 MOCs for operand row copy are counted in total MUL MOCs, and 1 MOC for MAC result store is counted in total ACC MOCs. Thus, by performing 16 MAC operations in parallel, *ATRIA* achieves shorter per-MAC latency.

*Second*, *ATRIA* can better hide the latency for PC operations, compared to SCOPE. This is because, SCOPE utilizes full adder-based PC operations that need to be performed inside PEs. Therefore, despite using the as-late-as-possible (ALAP) scheduling algorithm, PC operations in SCOPE inevitably stall the PEs. In contrast, *ATRIA* offloads PC operations to dedicated serial counters (operating at 2GHz) per PE (Sections III.B and III.C). As a result, *ATRIA* does not need to stall PEs for PC operations, enabling itself to better hide PC latency. Therefore, although *ATRIA* yields higher latency per PC operation than SCOPE (Table 3), *ATRIA* efficiently hides this higher latency, not letting it affect the performance.

*Third*, *ATRIA* exhibits smaller bottleneck ratio compared to SCOPE and DRISA (see Fig. 6(d) in Section IV.D). Bottleneck ratio is defined in Section IV.A. *ATRIA* achieves lower bottleneck ratio, because the use of massively large number of PEs in SCOPE and DRISA results in unavoidable inter-PE communication latency, a substantial portion of which remains on the critical processing path because of the inherently limited parallelism available for such inter-PE communications. In contrast, *ATRIA* is better at hiding the inter-PE communication latency, due to its smaller number of PEs and its LISA [26] substrate-based implementation of intra-bank, inter-bank, and inter-PE data communications (Section III.C).

**Table 3. Comparison of various accelerators with *ATRIA*, in terms of number of PEs (#PEs) and latency of MUL, ACC, MAC, binary to stochastic conversion (B-to-S), and pop count (PC) operations.**

| Various Accelerators | Latency Values | | | | | | #PEs |
|---|---|---|---|---|---|---|---|
| | MUL #MOCs | ACC #MOCs | MOC (ns) | MAC (ns) | B-to-S (ns) | PC (ns) | |
| DRISA-3T1C [1] | 200 | 11 | 8 | 1768 | - | - | 32768 |
| DRISA-1T1C-NOR [1] | 200 | 22 | 10 | 2110 | - | - | 16384 |
| LACC [3] | 1 | 10 | 21 | 231 | - | - | 16384 |
| SCOPE-Vanilla [2] | 3 | 4 | 8 | 56 | 1 | 176 | 65536 |
| SCOPE-H2D [2] | 21 | 4 | 8 | 200 | 1 | 176 | 65536 |
| *ATRIA* | 3/16 | 2/16 | 17 | 5.25 | 1 | 256 | 4098 |

## D. CNN Inference Performance Results

We evaluate the preformance of *ATRIA* and compare it with the following in-DRAM CNN accelerators from prior work: DRISA-3T1C [1], DRISA-1T1C-NOR [1], SCOPE-Vanilla [2], SCOPE-H2D [2], and LACC [3]. We consider four CNNs: VGG16 [14], Alexnet [13], ResNet_50 [16], GoogleNet [15], with the *ImageNet* dataset. Using the setup described in Section IV.A, we evaluated latency, FPS, FPS/W/mm$^2$, and bottleneck ratio, for batch size of 1 and 64.

Fig. 6(a) shows efficiency (FPS/W/mm$^2$) results. For batch size 1, *ATRIA* is 18×, 64×, 98× and 50× more efficient than DRISA-1T1C-NOR, DRISA-3T1C, SCOPE-Vanilla, and SCOPE-H2D, respectively, on average across CNNs. However, *ATRIA* is 15% less efficient than LACC, due to the LACC's lower area (Section III.D). Nevertheless, for batch size 64, *ATRIA* is more efficient than LACC as well. *ATRIA* is 136×, 522×, 3.4×, 71×, and 95× more efficient than DRISA-1T1C-NOR, DRISA-3T1C, LACC, SCOPE-Vanilla, and SCOPE-H2D, respectively, on average across CNNs. In general, *ATRIA* is more efficient due to the following two reasons: (i) better FPS due to lower per-MAC letency (Table 3), (ii) a reasonable average power consumption of 23.4W.

Fig. 6(b) shows CNN processing latency results normalized w.r.t. *ATRIA*. For batch size 1, *ATRIA* achieves 7.4×, 18×, 3.3×, 6.5×, and 4.4× lower latency than DRISA-1T1C-NOR, DRISA-3T1C, LACC, SCOPE-Vanilla, and SCOPE-H2D, respectively, on average across CNNs. Similarly, for batch size 64, *ATRIA* achieves 44×, 107×, 10×, 1.2×, and 2.6× lower latency than DRISA-1T1C-NOR, DRISA-3T1C, LACC, SCOPE-Vanilla, and SCOPE-H2D, respectively, on average across CNNs. *ATRIA* achives lower CNN processing latency because of its lower per-MAC latency and its ability of efficiently hiding its higher S-to-B conversion latency. Moreover, DRISA-1T1C-NOR, DRISA-3T1C, LACC, SCOPE-Vanilla, SCOPE-H2D, and *ATRIA* achieve 60×, 59×, 30×, 2×, 6×, and 10× higher latency for batch size 64 than batch size 1. This is because the higher parallelism of SCOPE variants (more #PEs in Table 3) allow them to process larger batch size with only a marginal increase in latency, by more efficiently distributing the batch processing across multiple PEs than any other accelerator.

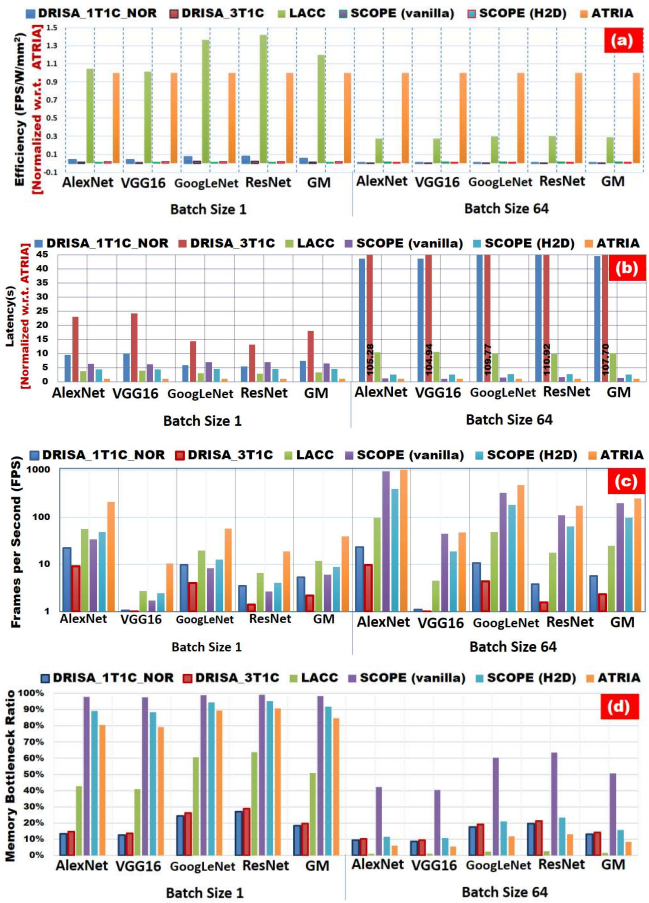

Fig. 6. (a) Efficiency (FPS/W/mm²), (b) latency, (c) throughput (FPS), and (d) memory bottleneck ratio (MBR) results for various in-DRAM accelerators across CNNs. GM means geometric mean.

Fig. 6(c) shows FPS results. For batch size 1, *ATRIA* has on average 7.4×, 18×, 3.3×, 6.5×, and 4.4× higher FPS than DRISA-1T1C-NOR, DRISA-3T1C, LACC, SCOPE-Vanilla, and SCOPE-H2D, respectively. For batch size 64, *ATRIA* has on average 44×, 107×, 10×, 1.2×, and 2.6× higher FPS than DRISA-1T1C-NOR, DRISA-3T1C, LACC, SCOPE-Vanilla, and SCOPE-H2D, respectively. *ATRIA* has higher FPS due to the combined effects of lower per-MAC latency and lower memory bottleneck ratio (Section IV.C), as discussed next.

Finally, Fig. 6(d) gives memory bottleneck ratio (MBR) results. MBR for all accelerators reduces for batch size 64 than batch size 1 because increasing batch size to 64 does not substantially increase the stall time for weighting parameter accesses, but doing so increases CNN processing time due to the required time-sharing of resources across multiple batch inputs, resulting in lower MBR. For batch size 64, *ATRIA* has lower MBR than all other accelerators, except for LACC. LACC has only 1% MBR for batch size 64, which corroborates with the results from [3]. This is because the kernel mapping algorithm used in LACC enables better resource utilization. SCOPE variants have the highest MBR for both batch sizes because in SCOPE the latency for S-to-B conversions come in the critical path (Section IV.C). In contrast, *ATRIA* is able to better hide this latency to achieve lower MBR.

## V. CONCLUSIONS

In this paper, we presented an energy-efficient and high-throughput CNN accelerator called *ATRIA*, which utilizes the novel concept of bit-parallel stochastic arithmetic to achieve ultra-low latency for multiply-accumulate (MAC) operations. We mapped four benchmark CNNs on *ATRIA* to compare its performance with five state-of-the-art in-DRAM accelerators from prior work. The results of our analysis show that *ATRIA* exhibits only 3.5% drop in CNN inference accuracy and still achieves improvements of up to 3.2× in frames-per-second (FPS) and up to 10× in efficiency (FPS/W/mm²), compared to the best-performing in-DRAM accelerator LACC from prior work. These results corroborate the excellent capabilities of *ATRIA* for accelerating the inference tasks of deep CNNs.

REFERENCES

[1] S. Li et al., "*Drisa: A dram-based reconfigurable in-situ accelerator,*" IEEE MICRO, pp. 288-301, 2017.
[2] S. Li et al., "*Scope: A stochastic computing engine for dram based in-situ accelerator,*" IEEE MICRO, pp. 696-709, 2018.
[3] Q. Deng et al., "*LAcc: Exploiting lookup Table-based fast and accurate vector multiplication in DRAM-based CNN accelerator,*" DAC, 2019.
[4] A. Ren et al., "*SC-dcnn: Highly-scalable deep convolutional neural network using stochastic computing,*" ACM ASPLOS, 2017.
[5] B. Jacob et. al., " *Memory systems: cache, DRAM, disk,*" Morgan Kaufmann, 2010.
[6] A. Alaghi et al., "*Survey of stochastic computing,*" IEEE TECS, pp.1-19, 2013.
[7] V. Sze et  al., "*Efficient processing of deep neural networks: A tutorial and survey,*" IEEE JPROC, pp.2295-2329, 2017.
[8] L. N. Smith, "*A disciplined approach to neural network hyper-parameters: Part 1--learning rate, batch size, momentum, and weight decay,*" arXiv preprint, 2018.
[9] Y. Chen et al., "*Dadiannao: A machine-learning supercomputer,*" IEEE MICRO, pp. 609-622, 2014.
[10] A. Shafiee et al., "*ISAAC: A convolutional neural network accelerator with in-situ analog arithmetic in crossbars,*" ACM ISCA, 44(3), pp.14-26, 2016.
[11] V. Seshadri et al., "*Ambit: In-memory accelerator for bulk bitwise operations using commodity DRAM technology,*" IEEE MICRO, pp. 273-287, 2017.
[12] C. Eckert et al., "*Neural cache: Bit-serial in-cache acceleration of deep neural networks,* " ACM ISCA, pp. 383-396, 2018.
[13] A. Krizhevsky et al., "*Imagenet classification with deep convolutional neural networks,*" Advances in neural information processing systems, *25*, pp.1097-1105, 2012.
[14] K. Simonyan et al., "*Very deep convolutional networks for large-scale image recognition*," arXiv preprint arXiv:1409.1556, 2014.
[15] C. Szegedy et al., " *Going deeper with convolutions,*" IEEE CVPR, pp. 1-9, 2015.
[16] K. He et Al., "*Deep residual learning for image recognition,*" IEEE CVPR, pp. 770-778, 2016.
[17] Balasubramonian et al., "*CACTI 7: New tools for interconnect exploration in innovative off-chip memories,*" TACO, *14*(2), pp.1-25, 2017.
[18] Q. Lou et al., "*A mixed signal architecture for convolutional neural networks,*" ACM JETC, *15*(2), pp.1-26, 2019.
[19] B. Zamanlooy et al., "*Efficient VLSI implementation of neural networks with hyperbolic tangent activation function,*" IEEE TVLSI, *22*(1), pp.39-48, 2013.
[20] R. Zunino et al., "*Analog implementation of the softmax function,*" IEEE ISCS, vol. 2, pp. II-II, 2002.
[21] Deng et al., "*DrAcc: a DRAM based accelerator for accurate CNN inference,*" IEEE DAC, pp. 1-6, 2018.
[22] C. Ping et al., "*Prime: A novel processing-in-memory architecture for neural network computation in reram-based main memory,*"ACM ISCA, *44*(3), pp.27-39, 2016.
[23] S. Xiaoyu et al., "*XNOR-RRAM: A scalable and parallel resistive synaptic architecture for binary neural networks,*" IEEE DATE, pp. 1423-1428, 2018.
[24] S. Han et al., "*EIE: Efficient inference engine on compressed deep neural network,*." IEEE ISCA, *44*(3), pp.243-254, 2016.
[25] V. Seshadri et al., "*RowClone: Fast and energy-efficient in-DRAM bulk data copy and initialization,*" IEEE MICRO, pp. 185-197, 2013.
[26] K. Chang et al. "*Low-cost inter-linked subarrays (LISA): Enabling fast inter-subarray data movement in DRAM,*" IEEE HPCA, pp. 568-580, 2016.
[27] Z. Li et al., "*Dscnn: Hardware-oriented optimization for stochastic computing based deep convolutional neural networks,*" IEEE ICCD, pp. 678-681, 2016.
[28] I. Thakkar et al., "*3D-ProWiz: An energy-efficient and optically-interfaced 3D DRAM architecture with reduced data access overhead,*" IEEE TMSCS, 1(3), pp.168-184, 2015.